

\message
{JNL.TEX version 0.95 as of 5/13/90.  Using CM fonts.}

\catcode`@=11
\expandafter\ifx\csname inp@t\endcsname\relax\let\inp@t=\input
\def\input#1 {\expandafter\ifx\csname #1IsLoaded\endcsname\relax
\inp@t#1%
\expandafter\def\csname #1IsLoaded\endcsname{(#1 was previously loaded)}
\else\message{\csname #1IsLoaded\endcsname}\fi}\fi
\catcode`@=12

\font\twelverm=cmr12			\font\twelvei=cmmi12
\font\twelvesy=cmsy10 scaled 1200	\font\twelveex=cmex10 scaled 1200
\font\twelvebf=cmbx12			\font\twelvesl=cmsl12
\font\twelvett=cmtt12			\font\twelveit=cmti12
\font\twelvesc=cmcsc10 scaled 1200	\font\twelvesf=cmssdc10 scaled 1200
                     
\font\twelvemib=cmmib10 scaled 1200
\font\tenmib=cmmib10
\font\eightmib=cmmib10 scaled 800

\skewchar\twelvei='177			\skewchar\twelvesy='60
\skewchar\twelvemib='177

\newfam\mibfam

\def\twelvepoint{\normalbaselineskip=12.4pt plus 0.1pt minus 0.1pt
  \abovedisplayskip 12.4pt plus 3pt minus 9pt
  \belowdisplayskip 12.4pt plus 3pt minus 9pt
  \abovedisplayshortskip 0pt plus 3pt
  \belowdisplayshortskip 7.2pt plus 3pt minus 4pt
  \smallskipamount=3.6pt plus1.2pt minus1.2pt
  \medskipamount=7.2pt plus2.4pt minus2.4pt
  \bigskipamount=14.4pt plus4.8pt minus4.8pt
  \def\rm{\fam0\twelverm}          \def\it{\fam\itfam\twelveit}%
  \def\sl{\fam\slfam\twelvesl}     \def\bf{\fam\bffam\twelvebf}%
  \def\mit{\fam 1}                 \def\cal{\fam 2}%
  \def\sc{\twelvesc}		   \def\tt{\twelvett}%
  \def\sf{\twelvesf}               \def\mib{\fam\mibfam\twelvemib}%
  \textfont0=\twelverm   \scriptfont0=\tenrm   \scriptscriptfont0=\sevenrm
  \textfont1=\twelvei    \scriptfont1=\teni    \scriptscriptfont1=\seveni
  \textfont2=\twelvesy   \scriptfont2=\tensy   \scriptscriptfont2=\sevensy
  \textfont3=\twelveex   \scriptfont3=\twelveex\scriptscriptfont3=\twelveex
  \textfont\itfam=\twelveit
  \textfont\slfam=\twelvesl
  \textfont\bffam=\twelvebf \scriptfont\bffam=\tenbf
                            \scriptscriptfont\bffam=\sevenbf
  \textfont\mibfam=\twelvemib \scriptfont\mibfam=\tenmib
                              \scriptscriptfont\mibfam=\eightmib
  \normalbaselines\rm}


\mathchardef\alpha="710B
\mathchardef\beta="710C
\mathchardef\gamma="710D
\mathchardef\delta="710E
\mathchardef\epsilon="710F
\mathchardef\zeta="7110
\mathchardef\eta="7111
\mathchardef\theta="7112
\mathchardef\iota="7113
\mathchardef\kappa="7114
\mathchardef\lambda="7115
\mathchardef\mu="7116
\mathchardef\nu="7117
\mathchardef\xi="7118
\mathchardef\pi="7119
\mathchardef\rho="711A
\mathchardef\sigma="711B
\mathchardef\tau="711C
\mathchardef\phi="711E
\mathchardef\chi="711F
\mathchardef\psi="7120
\mathchardef\omega="7121
\mathchardef\varepsilon="7122
\mathchardef\vartheta="7123
\mathchardef\varpi="7124
\mathchardef\varrho="7125
\mathchardef\varsigma="7126
\mathchardef\varphi="7127


\def\beginlinemode{\endmode
  \begingroup\parskip=0pt \obeylines\def\\{\par}\def\endmode{\par\endgroup}}
\def\beginparmode{\endmode
  \begingroup \def\endmode{\par\endgroup}}
\let\endmode=\par
{\obeylines\gdef\
{}}
\def\singlespace{\baselineskip=\normalbaselineskip}

\def\oneandahalfspace{\baselineskip=\normalbaselineskip
  \multiply\baselineskip by 3 \divide\baselineskip by 2}
\def\doublespace{\baselineskip=\normalbaselineskip \multiply\baselineskip by 2}

\newcount\firstpageno
\firstpageno=2
\footline={\ifnum\pageno<\firstpageno{\hfil}\else{\hfil\twelverm\folio\hfil}\fi}
\def\toppageno{\global\footline={\hfil}\global\headline
  ={\ifnum\pageno<\firstpageno{\hfil}\else{\hfil\twelverm\folio\hfil}\fi}}
\let\rawfootnote=\footnote		
\def\footnote#1#2{{\rm\singlespace\parindent=0pt\parskip=0pt
  \rawfootnote{#1}{#2\hfill\vrule height 0pt depth 6pt width 0pt}}}
\def\raggedcenter{\leftskip=4em plus 12em \rightskip=\leftskip
  \parindent=0pt \parfillskip=0pt \spaceskip=.3333em \xspaceskip=.5em
  \pretolerance=9999 \tolerance=9999
  \hyphenpenalty=9999 \exhyphenpenalty=9999 }
\def\dateline{\rightline{\ifcase\month\or
  January\or February\or March\or April\or May\or June\or
  July\or August\or September\or October\or November\or December\fi
  \space\number\year}}
\def\received{\vskip 3pt plus 0.2fill
 \centerline{\sl (Received\space\ifcase\month\or
  January\or February\or March\or April\or May\or June\or
  July\or August\or September\or October\or November\or December\fi
  \qquad, \number\year)}}


\hsize=6.5truein
\hoffset=0pt
\vsize=8.9truein
\voffset=0pt
\parskip=\medskipamount
\def\\{\cr}
\twelvepoint		
\doublespace		
\overfullrule=0pt	




\def\title			
  {\null\vskip 3pt plus 0.2fill
   \beginlinemode \doublespace \raggedcenter \bf}

\def\author			
  {\vskip 3pt plus 0.2fill \beginlinemode
   \singlespace \raggedcenter\sc}

\def\affil			
  {\vskip 3pt plus 0.1fill \beginlinemode
   \oneandahalfspace \raggedcenter \sl}

\def\abstract			
  {\vskip 3pt plus 0.3fill \beginparmode
   \oneandahalfspace ABSTRACT: }

\def\endtitlepage		
  {\endpage			
   \body}

\def\body			
  {\beginparmode}		

\def\head#1{			
  \goodbreak\vskip 0.5truein	
  {\immediate\write16{#1}
   \raggedcenter \uppercase{#1}\par}
   \nobreak\vskip 0.25truein\nobreak}

\def\itemitemitem{\par\indent\indent \hangindent3\parindent \textindent}
\def\itemitemitemitem{\par\indent\indent\indent \hangindent4\parindent
\textindent}
\def\beginitems{\par\medskip\bgroup
  \def\i##1 {\par\noindent\llap{##1\enspace}\ignorespaces}%
  \def\ii##1 {\item{##1}}%
  \def\iii##1 {\itemitem{##1}}%
  \def\iiii##1 {\itemitemitem{##1}}%
  \def\iiiii##1 {\itemitemitemitem{##1}}
  \leftskip=36pt\parskip=0pt}\def\enditems{\par\egroup}

\def\makefigure#1{\parindent=36pt\item{}Figure #1}

\def\figure#1 (#2) #3\par{\goodbreak\midinsert
\vskip#2
\bgroup\makefigure{#1} #3\par\egroup\endinsert}

\def\beneathrel#1\under#2{\mathrel{\mathop{#2}\limits_{#1}}}

\def\refto#1{$^{#1}$}		

\def\references			
  {\head{References}		
   \beginparmode
   \frenchspacing \parindent=0pt \leftskip=1truecm
   \parskip=8pt plus 3pt \everypar{\hangindent=\parindent}}

\gdef\refis#1{\item{#1.\ }}			

\gdef\journal#1, #2, #3, 1#4#5#6{		
    {\sl #1~}{\bf #2}, #3 (1#4#5#6)}		

\def\pr{\journal Phys. Rev., }

\def\prl{\journal Phys. Rev. Lett., }

\def\np{\journal Nucl. Phys., }

\def\endreferences{\body}

\def\figurecaptions		
  {\endpage
   \beginparmode
   \head{Figure Captions}
}

\def\endpage			
  {\vfill\eject}

\def\endpaper			
  {\endmode\vfill\supereject}


\def\heading				
  {\vskip 0.5truein plus 0.1truein	
   \beginparmode \def\\{\par} \parskip=0pt \singlespace \raggedcenter}

\def\subheading				
  {\vskip 0.25truein plus 0.1truein	
   \beginlinemode \singlespace \parskip=0pt \def\\{\par}\raggedcenter}

\def\tag#1$${\eqno(#1)$$}

\def\align#1$${\eqalign{#1}$$}

\def\aligntag#1$${\gdef\tag##1\\{&(##1)\cr}\eqalignno{#1\\}$$
  \gdef\tag##1$${\eqno(##1)$$}}

\def\overset #1\to#2{{\mathop{#2}\limits^{#1}}}
\def\underset#1\to#2{{\let\next=#1\mathpalette\undersetpalette#2}}
\def\undersetpalette#1#2{\vtop{\baselineskip0pt
\ialign{$\mathsurround=0pt #1\hfil##\hfil$\crcr#2\crcr\next\crcr}}}


\def\ref#1{Ref.~#1}			
\def\Ref#1{Ref.~#1}			
\def\[#1]{[\cite{#1}]}
\def\cite#1{{#1}}
\def\(#1){(\call{#1})}
\def\call#1{{#1}}
\def\taghead#1{}
\def\frac#1#2{{#1 \over #2}}

\def\12{{1\over2}}

\def\sla{\raise.15ex\hbox{$/$}\kern-.57em}
\def\leaderfill{\leaders\hbox to 1em{\hss.\hss}\hfill}
\def\twiddle{\lower.9ex\rlap{$\kern-.1em\scriptstyle\sim$}}
\def\bigtwiddle{\lower1.ex\rlap{$\sim$}}
\def\gtwid{\mathrel{\raise.3ex\hbox{$>$\kern-.75em\lower1ex\hbox{$\sim$}}}}
\def\ltwid{\mathrel{\raise.3ex\hbox{$<$\kern-.75em\lower1ex\hbox{$\sim$}}}}
\def\square{\kern1pt\vbox{\hrule height 1.2pt\hbox{\vrule width 1.2pt\hskip 3pt
   \vbox{\vskip 6pt}\hskip 3pt\vrule width 0.6pt}\hrule height 0.6pt}\kern1pt}
\def\tdot#1{\mathord{\mathop{#1}\limits^{\kern2pt\ldots}}}
\def\happyface{%
$\bigcirc\rlap{\lower0.3ex\hbox{$\kern-0.85em\scriptscriptstyle\smile$}%
\raise0.4ex\hbox{$\kern-0.6em\scriptstyle\cdot\cdot$}}$}
\def\sadface{%
$\bigcirc\rlap{\lower0.25ex\hbox{$\kern-0.85em\scriptscriptstyle\frown$}%
\raise0.43ex\hbox{$\kern-0.6em\scriptstyle\cdot\cdot$}}$}

\def\pmb#1{\setbox0=\hbox{#1}%
  \kern-.025em\copy0\kern-\wd0
  \kern  .05em\copy0\kern-\wd0
  \kern-.025em\raise.0433em\box0 }


\input jnl
\input  defs.tex
\input reforder
\input eqnorder
\input tables

\rightline{NSF-ITP-94-75}

\title Lattices of Matrices*

\vskip .2in

\author E. Br\'ezin$^1$ and A. Zee$^2$

\affil
$^1$Laboratoire de Physique Th\'eorique
\'Ecole Normale Sup\'erieure
24 Rue Lhomond
75231 Paris, France

\affil
$^2$ Institute for Theoretical Physics
University of California
Santa Barbara, CA 93106-4030 USA

\vskip .2in
*Revised version

\vskip .2in

\abstract
{We study a new class of matrix models, formulated on a lattice. On each
site are $N$ states with random energies governed by a Gaussian random
matrix Hamiltonian. The states on different sites are coupled randomly. We
calculate the
density of and correlation between the eigenvalues of the total Hamiltonian
in the large $N$ limit.
We
find that this correlation exhibits the same type of universal behavior we
discovered recently. Several derivations of this result are given. This class
of random matrices allows us to model the
transition between the ``localized" and ``extended" regimes within the
limited context of random matrix theory.}

\vskip .5in

We formulate and study a class of matrix models defined in the simplest
version by the random Hamiltonian
$$
H = \pmatrix{H_1 & V\cr V^\dagger& H_2\cr}
\eqno(hamiltonian)
$$
taken from the Gaussian distribution
$$
P(H)={1 \over Z}{e^{{-N}
tr ({1\over 2}(m^2_1 H^2_1 + m^2_2 H^2_2) +M^2 V^{\dagger} V})
}.\eqno(distribution)
$$
Here $H_{1,2}$, and $V$ denote $N$ by $N$ matrices, and $H$ a $2N$
by
$2N$ matrix, with $H$ and $H_{1,2}$ hermitean. The parameters
$m^2_{1,2}$
and
$M^2$ are both treated as order $N^0$ in the large $N$ limit.

A Hamiltonian of the form in \(hamiltonian) with $V$ equal to zero would
naturally arise when a conserved symmetry prevents two sectors in the
Hilbert space
of states to mix. The off-diagonal coupling $V$ would then represent
symmetry-breaking effects.

It is natural to generalize our model immediately to contain $C$ sectors
which mix due to some symmetry-breaking effects. The Hamiltonian $H$
in
\(hamiltonian) is generalized from a 2 by 2 block matrix to a $C$ by $C$
block matrix. Each block is an $N$
by
$N$ matrix, which we will denote by
$H_{\alpha\beta}$ where $\alpha$, $\beta$, ... run from 1 to $C$.
The distribution in \(distribution) is generalized to
$$
P(H)={1\over Z}e^{{-N}
 \sum_{\alpha, \beta} {{1\over 2}  \cM_{\alpha\beta}^2}
tr(H_{\alpha\beta}H_{\beta\alpha})}\eqno(general)
$$
Here $\cM^2$ represents a $C$ by $C$ real symmetric matrix. The
distribution considered in \(distribution) corresponds to the case
$C=2$ with $ \cM^2_{11} = m^2_1$, $ \cM^2_{22} = m^2_2$, and
$ \cM^2_{12}= \cM^2_{21}=M^2$.

It is natural also to think of the $C$ sectors as $C$ sites on a lattice. On
site $\alpha$ live $N$ states with energies
determined by $H_{\alpha\alpha}$. The states on sites $\alpha$ and $\beta$
are
coupled by $H_{\alpha\beta}$. The matrix $\cM^2$ determines the
``connectivity"
of the lattice. The standard matrix model corresponds to the
case of a lattice with one site. With this lattice interpretation, this model
is
essentially the same as the model proposed and studied by
Wegner\refto{wegner} some years ago. Our results overlap those of
Wegner; nevertheless it may be fruitful to approach the model from a
somewhat different point of view and within the context of our recent
study of universal correlation in matrix models.\refto{BZ1, BZ2, BZ3}

\refis{wegner} F. J. Wegner, \pr B19, 783, 1979; Z. f. Physik B49, 297,
1983.

This class of models does not appear
to
be analyzable by the standard method of orthogonal polynomials. On the
other hands, the density and correlation in this class of models can be
readily obtained using the diagrammatic approach recently developed by
us.\refto{BZ3} The counting and summing of diagrams provides a
somewhat amusing
exercise in perturbative field theory.

We note in passing how how we  are led to consider
such
a model. In localization theory it is well known that the distribution of
spacing
between neighboring energy levels changes character as one varies the
energy from the localized to the extended regime. The physics behind this
change is clear. Extended wave functions overlap in space, and any
perturbation would have a significant matrix element between two wave
functions close together in energy. The familiar phenomenon of level
repulsion as described by second order pertubation theory tells us that the
likelihood of having neighboring energy levels separated by an energy $s$
becomes vanishingly small as $s$ goes to zero. This simple argument,
essentially
due to Wigner\refto{wig57b} and to Landau and Smorodinsky,\refto{ls} is
known as the Wigner surmise. We expect the probability $p(s)$ of a
spacing $s$ between neighboring energy levels to rise from zero for
$s=0$, reach a peak, and then to decrease rapidly\refto{MEH}. The curve
traced out by $p(s)$ is sometimes referred\refto{econo} to informally as
``Wigner's
head." On the other hand, localized wave functions have little overlap with
each other, and thus do not repel each other significantly. In the localized
regime $p(s)$ simply decreases from some finite value as $s$ increases
from zero, following a Poisson distribution. The curve traced out by
$p(s)$ is known informally as ``Poisson's tail."

Recently, there have been some attempts\refto{mutt, neu} to incorporate
this transition from the ``Wigner regime" to the ``Poisson regime" in the
context of the
theory of random matrices.\refto{WIG, MEH, POR} In matrix models the
concept of space does not
appear of course; nevertheless one can ask whether the distribution of
eigenvalues, as manifested in the density of, and the correlation between,
eigenvalues may exhibit the type of transition described above for
localization theory. In particular, Moshe, Neuberger, and
Shapiro\refto{neu} proposed a new class of matrix models with a
parameter they called $b$ such that as $b$ varies the distribution of
eigenvalues
goes from a Wigner to a quasi-Poisson distribution. We found their model
somewhat unsatisfactory in that the parameter $b$ has to be order $N^2$
larger than the other parameters in the model, where $N$ as usual in
discussions of matrix models denotes the size of the matrices. This actually
reflects the universality of the Wigner distribution, that is, roughly
speaking, its
tendency to resist change. We feel that the model studied here does not
suffer from this difficulty.

Take the $C=2$ case as an example. We note that at the special
value $M^2 = m^2_1=m^2_2 =2 m^2_{eff}$ (which we will refer as the
Wigner point), the distribution \(distribution)
collapses
to
$$
P(H)={1\over Z}e^{-2N
tr {{1\over 2}m^2_{eff}} H^2}
\eqno(onematrix)
$$
the standard Gaussian distribution for $2N$ by $2N$ matrices and we
should
recover the results of \Ref{BZ3} as a check. For $M^2 = \infty$, the
sectors 1 and 2 clearly decouples, and the spacing
distribution is trivially Poissonian in the sense described above. We will
refer to this as the Poisson point.
For $M^2 < \infty$, the off-diagonal perturbation $V$ generates level
repulsion, and the spacing distribution should be Wignerian.

We now calculate the density of eigenvalues and the correlation between
eigenvalues using a diagrammatic method we developed
recently.\refto{BZ3} As usual, these quantities can be obtained from the
Green's functions
$$
\eqalignno{
G(z)&\equiv\vev{{1\over CN}tr{1\over z-
H}}&(1.3)\cr
G(z,w)_c
&\equiv\vev{{1\over CN}tr{1\over z-H}{1\over
CN}tr{1\over
w-H} }_c&
(1.4)
\cr}
$$
where
$
\vev{0 (H)} \equiv \int dH\  0(H)P(H)
$
and the subscript $c$ indicates the connected Green's
function.
The density of eigenvalues is then given by
$$
\rho(\mu)=\vev{{1\over N}tr\de(\mu-H)}={-
{1\over\pi}}{\rm Im}
 G(\mu+i\eps)
\eqno(1.6)
$$
and the correlation between eigenvalues, by
$$
\eqalign{
\rho_c(\mu,\nu)&=\vev{{1\over N}tr\de(\mu-
H){1\over
N}tr\de(\nu-
H)}_c\cr
&=(-1 /4\pi^2)(G_c(++)+G_c(--)-G_c(+-)-G_c(-
+))\cr}\eqno(1.7)
$$
with the obvious notation
$
G_c(\pm,\pm)\equiv G_c(\mu\pm i\eps,\nu\pm
i\de)
$
(signs uncorrelated).

The Feynman diagram expansion corresponds to an expansion in powers of
$1/z$ and $1/w$. Let us illustrate with the correlation function and expand
$$
(CN)^2 G_c(z,w)=\sum^\infty_{m=0}\sum^\infty_{n=0}{1\over
z^{m+1}w^{n+1}}
\vev{trH^m   trH^n}_c     \eqno(expand)
$$
The calculation proceeds along the same line as in \Ref{BZ3} suitably
generalized to include the block indices $\alpha$, $\beta$,...
Diagrammatically, we may borrow the
terminology of large N
QCD\refto{thoo}
and describe the expression for
 $G_c(z,w)$ as two separate quark loops,
of type $z$ and type $w$ respectively, interacting by
emitting
and absorbing gluons. Now each gluon line also carries two block indices
$\alpha\beta$.
Some readers may find it natural to think of the $C$ block indices as
representing
a second ``color," or ``technicolor" index. We take the limit $N$ large for
fixed $C$. With
the
Gaussian distribution in \(general)
we can readily
``Wick-contract" \(expand).
The gluon propagator is given by
$$
\vev{H_{\alpha\beta ij} H_{\gamma\epsilon kl}} =
{1\over N}\de_{il}\de_{jk}\de_{\alpha\epsilon}\de_{\gamma\beta}
\sigma_{\alpha\beta}\eqno(glueprop)
$$ where we have defined the $C$ by $C$ real symmetric matrix $\sigma$
by
$ \sigma_{\alpha\beta} = 1/\cM^2_{\alpha\beta}$.
 (Note that the matrix
$\sigma$ is not the inverse of the matrix $\cM^2$.) Thus, the gluon is
represented by a
double
line while a quark  is represented by a single line. This convention
greatly
facilitates counting the powers of $N$ as explained in \Ref{BZ3}.

Let us begin by calculating the propagator, which is easily seen
diagrammatically to be diagonal in block indices. Define $g_{\alpha}$ by
$$
G_{\alpha\beta, ij}=<({1\over z-H})_{\alpha\beta, ij}> \equiv
\de_{\alpha\beta}\de_{ij}g_\alpha
\eqno(self)
$$
In the large $N$ limit, the sum of the leading  planar diagrams
(``generalized rainbow" diagrams) is determined by the self-consistent
equation
$$
g_{\alpha}={1\over {z- \sum_{\beta} \sigma_{\alpha\beta} g_{\beta}}}
\eqno(propagator)
$$
For general $\cM^2$ this represents a system of coupled non-linear
equations to be solved for $g_{\alpha}$. It
may be interesting to study the solution numerically for arbitrary
$\cM^2$.

We are now ready to calculate $G_c(z,w)$. As in \Ref{BZ3} we begin by
ignoring in \(expand)
contractions
within
the same trace (in which case $m$ and $n$ are required to
be
equal). In the
large $N$ limit, the dominant graphs are given essentially by ``ladder
graphs"
(with one crossing). We now have a combinatorial problem which is most
easily solved by distorting the ladder graph in Fig. (1a) to the ``wheel"
graph
in Fig. (1b). (Here we admit distortions as long as they preserve the
combinatorial factor.) We see that each ``spoke" of the wheel is associated
with a factor
of $\sigma_{\alpha\beta}$. Thus, the graphs with $n$ loops
should
be
multiplied by the factor $tr \sigma^n$. We obtain
$$
(CN)^2 G_c(z,w) = \sum_{n=0}^\infty {n tr  \sigma^n
\over(zw)^{n+1}}
\eqno(ladder1)
$$

We next include Wick-contractions within the same trace
in
$\vev{trH^m
trH^n}$. Graphically these contractions
correspond to decorating the ladder graphs by vertex
and self
energy corrections. First, we have to
correct
the upper part of the graph in Fig. (1a) with vertex corrections, as shown
in
Fig. (2), and similarly
 for the lower part of the graph. These vertex corrections lead to a rather
complicated expression. For example,  the factor $tr
 \sigma^n$
in \(ladder1) for $n=3$ is to be replaced by the sum of terms, one of which
has the form
$$
\sum_{\alpha\rho\omega} \sigma_{\alpha\rho}\sigma_{\rho\omega}
\sigma_{\rho\alpha}\sum_{\beta}({1\over 1-
\sigma/z^2})_{\alpha\beta}\sum_{\gamma}({1\over 1-
 \sigma/w^2})_{\alpha\gamma}
\eqno(mess)
$$
and another with the factor $\sum_{\gamma}({1\over 1-
 \sigma/w^2})_{\alpha\gamma}$
 in the expression above replaced by
$\sum_{\gamma}$ $({1\over 1-
 \sigma/w^2})_{\omega\gamma}$. This clearly leads to a rather
unwieldy expression into which we still have to put in the self energy
corrections.

Clearly we should not expect to obtain a simple and closed expression for
arbitrary $\sigma$: every sector or every site would be unique in its local
properties. Instead, we made the rather mild and physically reasonable
assumption of homogeneity by supposing that
$$
\sum_{\beta}
\sigma_{\alpha\beta} = {\rm independent\ of\/}\  \alpha.
\eqno(condition)
$$
Every site (or sector, if the
reader prefers) is treated on the
same footing. In particular, this assumption holds if the lattice is
translation
invariant.

Quite remarkably, we now notice that this model may in fact be solved  on
any
lattice
in any dimension as long as the condition  \(condition) holds. Consider the
eigenvalue decomposition the real symmetric matrix $\sigma$, which  we
may identify as a ``hopping matrix",
$$
\sigma = \sum_k  |k> \epsilon_k <k|
\eqno(eigen)
$$
where $k$ takes on $C$ values. Denote the eigenvalues by
$\epsilon_k$. The condition \(condition) merely states
that
of the eigenvectors of $\sigma$ there is one (which we labeled as $|k=0>$)
with components all equal to $1/{\sqrt C}$.

We find immediately that in \(propagator)
 $g_{\alpha}$ is independent of $\alpha$
and equal to
$$
G(z)={{z-\sqrt {z^2 - 4\epsilon_0} }\over 2 \epsilon_0}
\eqno(prop)
$$
This has the same form as the propagator in the simple one
matrix model and gives immediately Wigner's
semi-circle law for the density of eigenvalues
$$
\rho (\mu) = {1\over 2\pi \epsilon_0} \sqrt{ 4\epsilon_0 - \mu^2}
\eqno(density)
$$

We now proceed to the two-point Green's function. The vertex correction
factor we encountered before now
simplifies drastically
$$
\sum_{\beta}({1\over 1-
\sigma/z^2})_{\alpha\beta} = ({1\over 1-
\epsilon_0/z^2})
\eqno(simple)
$$
to a factor independent of $\alpha$. Putting in vertex corrections just
amounts to multiplying by this factor and a corresponding one with $z$
replaced by $w$. The trace and sum in \(ladder1) can be performed
immediately
to give
$$
(CN)^2 G_c (z,w) = \sum_k {\epsilon_k \over (zw - \epsilon_k)^2}.
\eqno(ladder3)
$$
After putting in vertex and self-energy corrections, we find
$$
(CN)^2 G_c(z,w) =\Big( {G^2(z) \over 1- \epsilon_0 G^2(z)}\Big)
\Big({G^2(w) \over 1- \epsilon_0 G^2(w)}\Big)
\Big(\sum_k {\epsilon_k
\over
(1 - \epsilon_k G(z)G(w))^2}
\Big).
\eqno(central)
$$

Taking the appropriate absorptive parts we can calculate the connected
correlation
function $\rho_c(\mu,\nu)$. As in our earlier work we find it natural to
introduce angular variables defined by
$$
\sin \theta = {\mu \over {2\epsilon_0}}
\eqno(theta)
$$
and
$$
\sin \varphi ={ \nu \over {2\epsilon_0}}.
\eqno(varphi)
$$ The angles $\theta$ and $\varphi$ vary from $-\pi/2$ to $+\pi/2$ over
the range of the eigenvalues.
The connected correlation function is given by
$$
\eqalign{
\rho_c(\mu,\nu) &=-
{1\over16\pi^2N^2C^2\epsilon_0\cos\theta\cos\varphi} \cr
\times \sum_k\tau_k & \Big\{
{\tau_k+\cos(\theta+\varphi)\over(1+\tau_k\cos(\theta+\varphi))^2} +
{\tau_k-\cos(\theta-\varphi)\over(1-\tau_k\cos(\theta-\varphi))^2} \Big\}
}
\eqno(correlation')
$$
where we have defined
$$
\tau_k = {2\epsilon_0 \epsilon_k \over {\epsilon_0^2 + \epsilon_k^2}}.
\eqno(tauk)
$$
The expressions \(central) and \(correlation') represent the central results
of this paper. It is worth emphasizing that our results hold for any
lattice with any connectivity as long as the homogeneity condition
\(condition) is
satisfied.

As a simple special case, consider the simple case in which the matrix
$\cM^2$
is such that all its diagonal elements are equal to $m^2$ and all its
off-diagonal elements are equal to $M^2$. Every site on the lattice couples
to
every other site. The matrix $\sigma$ has  one eigenvalue
$\epsilon_0=1/{m_s^2}$ and
$C-1$
degenerate eigenvalues equal to $1/{m_d^2}$. Thus we have
$$
\eqalign{
\rho_c(\mu,\nu) &=- {m_s^2\over16\pi^2N^2C^2\cos\theta\cos\varphi}
\Big\{
{1\over1+\cos(\theta+\varphi)} + {1\over1-\cos(\theta-\varphi)} \cr
+(C-1) &\tau
{\tau+\cos(\theta+\varphi)\over(1+\tau\cos(\theta+\varphi))^2} +
(C-1) \tau {\tau-\cos(\theta-\varphi\over(1-\tau\cos(\theta-\varphi))^2}
\Big\}
}
\eqno(correlationspecial)$$
where
$$
\tau = {2 m_s^2 m_d^2 \over {m_s^4 + m_d^4}}.
\eqno(tau)
$$ Note that $|\tau| < 1$ and thus only the second term in
$\rho_c(\mu,\nu)$ is singular when
$\mu = \nu$. As we vary $M^2$, the correlation moves from a Poissonian
(``localized") regime to a Wignerian (``extended")
regime. In particular, at the Wigner point, $\tau = 0$ and we recover
trivially the result for the one matrix model first obtained in \Ref{BZ1}.
At the Poisson point, $\tau = 1$ we have a short distance (i.e., when
$\theta$ and $\varphi$ approach each other) singularity with a residue $(C-
1+1)/C^2$. (Recall from \Ref{BZ1} that this short distance singularity
occurs because we are dealing with the ``smoothed" correlation. The exact
correlation does not have this singularity. Nevertheless, this singularity
tells us about the short distance correlation between the density of
eigenvalues.)
Thus, the short distance singularity is reduced from the standard $C=1$
case and is completely suppressed in the large $C$ limit, namely in the
limit of many sectors forbidden to mix by many conservation laws. More
generally, for any $\tau$, it is sufficient to take $C$ large for this short
distance singularity to be suppressed.\refto{timereg}

For an arbitrary lattice, we
simply
find the eigenvalues of the hopping matrix $\sigma$. For example, for a
hypercubic lattice we have
$$
\epsilon_k = {1\over m^2}+{2\over M^2} \sum_a \cos k_a
\eqno(epsilon)
$$
where the sum over $a$ runs overs the dimension $D$ of the lattice. In the
expressions \(central) and \(correlation') the sum over $k$ runs as usual
over the Brillouin zone $k_a= 2\pi j_a/L_a,
j_a=0,1,2,...L_a-1$ with $L_1 L_2 ... L_D = C$. Note that to have nearest
neighbor
coupling the appropriate entries in the matrix  $\cM^2$ have to be taken to
infinity. This simply means that some of the matrices $H_{\alpha\beta}$ in
\(general) are to be set equal to zero.  It may be interesting to study
our result for a variety of lattices.
Also, one may consider the limit $C \rightarrow \infty$ in
which
case the sum over $k$ is replaced by an integral. While our results are
obtained in the context of random matrix theory, they may conceivably be
of relevance to some condensed matter systems.

 This correlation function, while it is not quite the same as the correlation
functions we\refto{BZ1,BZ2,BZ3} and others\refto{bee,eyn} have found
previously in a number
of different
situations, has the same general structure as these
other correlation functions.\refto{foot} This in itself is a statement of
universality.
Next, we may make a remark about going beyond Gaussian
distributions. In our previous work\refto{BZ2,BZ3}
we identified two classes of matrix models: the Wigner class
 and the trace class. Briefly, in the Wigner class  the
individual matrix element obeys a probability distribution.
(In the present context, the distributions for different $H_{\alpha\beta}$
may in general be different.) As was explained in \Ref{BZ2},
universality, that is, the independence of the density and the
correlation on these probability distributions, can be proved
almost immediately using our diagrammatic method. In contrast,
in the trace class, each of the $H_{\alpha\beta}$ obeys some
 probability distribution of the form
$$
P(H_{\alpha\beta}) \propto e^{- N tr V^{\alpha\beta} (H_{\alpha\beta})}
\eqno(traceclass)
$$
(no sum over repeated indices) where $V^{\alpha\beta}$
denotes a polynomial, different for different ${\alpha\beta}$.
In this case, we do not know how to prove universality, that is,
 the independence of the correlation function $\rho_c$ (when expressed
in terms of scaling variables) on $V^{\alpha\beta}$.  It would be
interesting to see numerically whether  universality indeed holds.

We would now like to present two other derivations of our central result.
Let us for notational simplicity  consider the
 case studied in \Ref{BZ3}, that of one Gaussian-random matrix,
which we will denote by $H$. In other words, we first consider $H$ for
$C=1$, a lattice with one point. We will first recover the results of
\Ref{BZ3} and then generalize these results to an arbitrary lattice.
Consider
the ``scattering amplitude"
$$
\eqalign{\Gamma_{ln}^{im}&\equiv
N\vev{({1\over{z-H}})^i_l({1\over{w-H}})^m_n}_c
\cr&=N(\vev{({1\over{z-H}})^i_l({1\over{w-H}})^m_n} -
\delta^i_l G(z) \delta^m_n G(w))
\cr&\equiv \delta^i_n \delta^m_l A + \delta^i_l  \delta^m_n B\cr}
\eqno(scattering)
$$ The two ``scalar" functions $A$ and $B$ depend on $z$ and $w$ of
course.
The Green's function is defined as usual by
$$
G(z)\equiv\vev{{1\over N}tr{1\over{z-H}}}
\eqno(green)
$$
we now calculate (repeated indices summed)
$$
\Gamma^{il}_{li} = N^2 A +NB
= N^2 [{{G(z)-G(w)}\over {w-z}} - G(z) G(w)]
\eqno(abc)
$$
This equation tells us that $A$ is of order $N^0$ while $B$ is of order
$1/N$. Furthermore, the connected two-point Green's function is given by
$$
N^2 G_c(z,w)\equiv\vev{tr{1\over{z-H}}tr{1\over{w-H}}}_c
=A + NB
\eqno(twopoint)
$$
Thus, if we know $G$ and $G_c$ we know $A$ and $B$ and hence the full
scattering amplitude $\Gamma_{ln}^{im}$.

For our lattice problem we have to generalize the definition above to
$$
\eqalign{\Gamma_{\beta l \epsilon n}^{\alpha i \gamma m}&\equiv
N\vev{({1\over{z-H}})^{\alpha i}_{\beta l}({1\over{w-H}})^{
\gamma m}_{\epsilon n}}_c\cr
&\equiv \delta^{\alpha}_{\epsilon} \delta^i_n
\delta^{\gamma}_{\beta}\delta^m_n A + \delta^{\alpha}_{\beta} \delta^i_l
\delta^{\gamma}_{\epsilon} \delta^m_n B\cr}
\eqno(latticescattering)
$$
All is as before except that now the scattering functions $A$ and $B$ can
depend on the site labels $\alpha$, $\beta$, $\gamma$ and $\epsilon$ or
technicolor indices because the probability distribution is not unitary
invariant in these indices. Note that nevertheless the kronecker delta
structure in \(latticescattering) holds. This can be seen graphically since a
line carrying color and technicolor going in must come out carrying the
same color and technicolor.

Now we note the remarkable fact that if we are interested in questions of
localization, namely what happens when the site $\alpha$ is equal to the site
$\epsilon$, and far away from the site $\beta$=$\gamma$, we don't need to
know anything about $B$. But $A$ is determined by the Green's function
$G(z)$ completely!

Recalling that for the Gaussian and single matrix case (see \(propagator)
and \(prop)) that $G(z)$ is the solution of the quadratic equation
$$
z=G+{1\over G}
\eqno(quadratic)
$$
we obtain immediately from \(abc) that
$$
A(z,w)=  {G^2(z) G^2(w) \over 1 - G(z) G(w)}
\eqno(a)
$$
One can see readily that for the case of a lattice of matrices considered here
we only have to modify this result to
$$
\eqalign{A^{\alpha}_{\beta}(z,w)&=  G^2(z) G^2(w) ({\sigma \over {1 -
\sigma
G(z) G(w)} })^{\alpha}_{\beta} \cr
&=G^2(z) G^2(w) \int {d^dk\over (2\pi)^d} {{\epsilon_k e^{i{\vec k}
\cdot
{(\vec \alpha - \vec \beta)}}} \over {1 - \epsilon_k G(z) G(w)} }\cr}
\eqno(diff)
$$
Note that as anticipated $A$ depends on $\alpha$ and $\beta$.

This result appears to agree with Wegner's result\refto{wegner}. We
would like to emphasize the simplicity and ease with which this result can
be obtained. For the sake of completeness, let us recall that the standard
localization formalism instructs us to set $z=E+i\omega$ and $w=E-
i\omega$ and then let the distance $r\equiv |\vec \alpha - \vec \beta|$ tend to
infinity. Since $G(E+i\omega)G(E-i\omega) \rightarrow 1$ as $\omega
\rightarrow 0$ and $\epsilon_k \rightarrow 1 - O(k^2)$ as $k \rightarrow
0$ for a simple rectilinear lattice, we see that $A \rightarrow 1/r^{d-2}$,
as is well known from the work of Wegner.

As a check, we may also easily calculate $A$ directly. From the structure
of the indices we see that $A$ is given by summing the ladder graphs in
Fig (3). Note that vertex corrections are suppressed. We merely have to
dress the quark propagators. We thus have
$$
A(z,w) = {1\over zw} \sum_{n=1} {1\over (zw)^n }
(\sigma^n)^{\alpha}_{\beta}
\eqno(ladderforA)
$$
After summing and dressing the quark propagators by replacing $1/z$ by
$G(z)$ and $1/w$ by $G(w)$ we obtain \(diff) immediately.

In this approach, in order to obtain our correlation function, we must
calculate $B$.
Thus, there is conservation of labor after all. While it is somewhat more
involved to calculate $B$ than to calculate $A$, it is still easy enough. The
relevant graphs are given in Fig. (4). Curiously, because of the structure of
indices only graphs with vertex corrections contribute to $B$. In
particular, the simple one-gluon exchange graph does not contribute. We
obtain (again for notational simplicity we do the $C=1$ case here)
$$
\eqalign{NB(z,w) &= {(\sum^{\infty}_{p=0}\sum^{\infty}_{q=0}{1\over
z^{2p}w^{2q}} - 1) }
\sum^{\infty}_{n=0}{1 \over (zw)^{n+2}}\cr
&\rightarrow { {G^2(z)  +G^2(w)-G^2(z) G^2(w)}\over {(1-G^2(z))(1-
G^2(w))  } }{G^2(z) G^2(w) \over 1 - G(z) G(w)}\cr}
\eqno(graphsforB)
$$
after dressing the quark propagators. Note the $(-1)$ inside the
parenthesis. We leave it to the reader to show that
when this somewhat strange looking expression is combined with $A$
according to \(twopoint) we recover our
result\refto{BZ1, BZ3} for the correlation function $G_c(z,w)$. It is a
straightforward
exercise to put in the $\sigma$ matrix to recover \(central).

We conclude by giving yet another derivation of our central result. Note
that \(1.4) may be written as
$$
\eqalign{G_c(z,  w) &\equiv\vev{{1\over CN}tr{1\over z-H}{1\over
CN}tr{1\over w-H} }_c \cr &
={\part\over\part z}{\part\over\part w}{\vev{{1\over CN}tr{\log (z-
H)}{{1\over CN}tr{\log (w-H)}}}} \cr}
\eqno(logform)
$$
Again, for notational simplicity, let us consider the $C=1$ case first.
Expanding the logarithms, we find
$$
G_c(z,w)={\part\over\part z}{\part\over\part w}
{\sum^{\infty}_{n=1}\sum^{\infty}_{m=1}
{1\over z^n w^m}{\vev{{1\over
Nn}tr H^n {1\over Nm}tr H^m}}}
\eqno(more)
$$
This is represented by the RwheelS graph of Fig (1b). We see immediately
that vertex corrections are suppressed in the large $N$ limit. Thus, we may
set $n=m$ in \(more) and since $\vev{tr H^n tr H^n} = n$ we can
immediately evaluate the sum to obtain
$$
N^2G(z,w)_c=- {\part\over\part z}{\part\over\part w}\log (1- {1\over
zw})
\eqno(naked)
$$
Dressing the quark propagators, we obtain immediately
$$
N^2G(z,w)_c= - {\part\over\part z}{\part\over\part w}\log (1-
G(z)G(w))
\eqno(dressed)
$$
This derivation is simpler than that given in \Ref{BZ3}. We also obtain a
more compact form as given in \Ref{bhz}. It is now simple to go to the
case of an arbitrary lattice. We now have $\vev{tr H^n tr H^n} = n tr
\sigma^n$ and thus we find
$$
N^2G(z,w)_c= - {\part\over\part z}{\part\over\part w}tr{\log (1- \sigma
G(z)G(w))}
= - \sum_k {\part\over\part z}{\part\over\part w}{\log (1- \epsilon_k
G(z)G(w))}
\eqno(latticelog)
$$
We leave it to the reader to check that we have indeed recovered our
previous result \(central).

\refis{bhz} E.Br\'ezin, S. Hikami, and A. Zee, Paris-Tokyo-Santa-Barbara
preprint, 1994.

Finally, we are tempted to indulge in a speculation. It has long been known
that matrix models are intimately related to the Calogero-Sutherland
model. Specifically, the Calogero-Sutherland model for a certain coupling
constant $\lambda$ equal to 1/2, 1, and 2 correspond respectively to matrix
models with real symmetric, hermitean, and quarternionic matrices.
Recently, Ha\refto{ha} was able to solve the Calogero-Sutherland model
for $\lambda$
equal to an arbitary rational number. The question naturally arises whether
or not for these values of the coupling constant the Calogero-Sutherland
model corresponds to matrix models. Is it possible that the class of matrix
models considered here may be the correspondents? And if not, it would be
interesting to ask whether the matrix models considered here corresponds
to some generalized Calogero-Sutherland models.

\head{Acknowledgement}

We thank the referee of an earlier version of this
paper for calling our attention to the work of Wegner. This work was
supported in part by the National Science
Foundation under Grant No. PHY89-04035 and the Institut
Universitaire de France. EB would like to acknowledge the Institute for
Theoretical Physics, Santa Barbara, where this work was initiated, for its
hospitality. AZ acknowledges the Institute for Advanced Study, Princeton,
and the \'Ecole Normale Sup\'erieure, Paris,
where part of this work was done, for their hospitality.

\head{Figure Captions}

Fig (1a) A crossed ladder graph that contributes to leading order in the
large $N$ limit; (1b) As far as the combinatorics is concerned, graphs of
the type drawn in (1a) are equivalent to the graphs of the type drawn in
(1b).

Fig (2) Vertex corrections to the upper part of the graph in (1a)

Fig (3) Graphs contributing to $A$ in the large $N$ limit.

Fig (4) Some graphs contributing to $B$; note that only graphs with vertex
corrections are to be included.

\references

\refis{BZ1} E. Br\'ezin and A. Zee, \np 402(FS), 613, 1993.

\refis{BZ2} E. Br\'ezin and A. Zee, \np 424(FS), 435, 1994.

\refis{BZ3} E. Br\'ezin  and A. Zee, \pr E49, 2588, 1994.

\refis{ha} Z. N. C. Ha, Princeton IAS preprint 1994

\refis{neu} M. Moshe, H. Neuberger, and B. Shapiro, Technion preprint
PH-12-94 cond-mat/9403085.

\refis{WIG} E. Wigner, {\sl Can.\ Math.\ Congr.\ Proc.\/}
p.174
(University of
Toronto Press) and other papers reprinted in Porter, op.
cit.

\refis{POR} C.E. Porter, {\it Statistical\ Theories\ of \
Spectra:\ \
Fluctuations\/}
(Academic Press, New York, 1965).

\refis{MEH} M.L. Mehta, {\it Random\ Matrices\/}
(Academic
Press, New
York,
1991) See figures (1.3-1.5).

\refis{econo}  E.N. Economou, private communication to A.Z.

\refis{ls} L. Landau and Ya. Smorodinski, Gorsar. Izd. Tex.-teo. Lit.,
Moscow (1955) p92-93.

\refis{wig57b} E. P. Wigner, Oak Ridge Natl.Lab. ORNL-2309 (1957) p.
59-70.

\refis{bee} C.W.J. Beenakker, Institut Lorentz preprint
(1993) cond-mat/9310010; for earlier work, see C.W.J.
Beenakker, \prl 70, 1155, 1993; \pr
B47, 15763, 1993.

\refis{eyn} B. Eynard, Saclay preprint to appear in {\sl Nucl. Phys.} FS,
1994.

\refis{mutt} K.A. Muttalib, Y. Chen, M.E.H. Ismail, and V.N. Nicopoulos,
\prl 71, 471, 1993.







\refis{thoo} G. 't Hooft, \np B72, 461, 1974.

\refis{foot} In particular, some of the terms in \(correlation') have the
same
form as the terms in equation (2.18) of the time dependent case studied in
\Ref{BZ3}, with the identification $\tau = 1/\cosh u $.

\refis{timereg} This is related to the remark in \Ref{BZ3} that time acts as
a regulator for the short distance singularity in the time dependent problem
studied there.

\endreferences

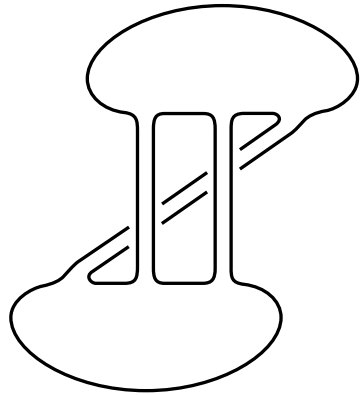
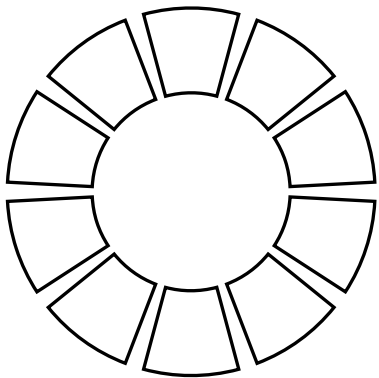
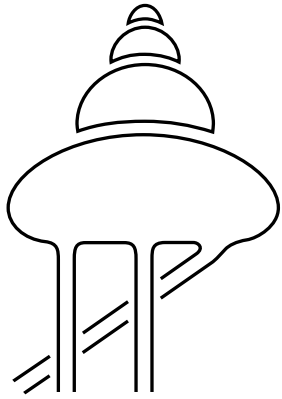
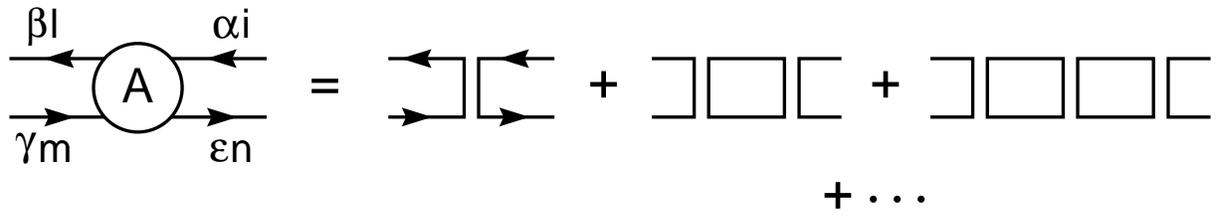
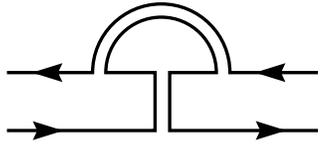
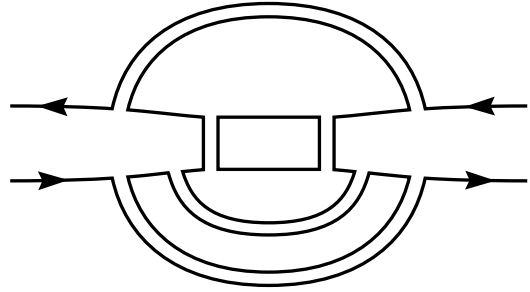
\end